\documentstyle[11pt]{article}

\font\rauth=cmcsc10
\newcommand{\ds}{\displaystyle}
\newcommand{\magsim}
{\ \lower2pt\hbox{$\sim $}\mkern-14mu \raise2pt\hbox{$>$}\ }
\newcommand{\minsim}
{\ \lower2pt\hbox{$\sim $}\mkern-14mu \raise2pt\hbox{$<$}\ }
\newcommand{\magmin}
{\ \lower3pt\hbox{$<$}\mkern-14mu \raise3pt\hbox{$>$}\ }
\newcommand{\um}[1]{\hspace{.15truecm}\mbox{\sl #1}}
\newcommand{\umtab}[1]{\mbox{\sl #1}}
\newcommand{\msun}{M_\odot}
\newcommand{\eq}[1]{equation~(#1)}
\newcommand{\al}{{\sl et al.\ }}
\newcommand{\paper}[5]
{\item {\rauth #1} #2, {\it #3}, {\bf #4}, #5.}
\newcommand{\iom}[4]
{\item {\rauth #1} #2, {\it #3}, {\bf #4}.}
\newcommand{\book}[4]
{\item {\rauth #1} #2, {\it #3}, (#4).}
\newcommand{\proc}[5]
{\item {\rauth #1} #2, {in \it #3}, {ed.\ #4} (#5).}

\title{\bf SINUSOIDAL GRAVITATIONAL WAVES FROM THE NUCLEI OF ACTIVE
GALAXIES}
\author{{\bf Giacomo Giampieri}\thanks{E-mail: gxg@grouch.jpl.nasa.gov}\\
\\{\it Jet Propulsion Laboratory}\\
{\it California Institute of Technology}\\
{\it MS 301-150 -- 4800 Oak Grove Drive, Pasadena, CA 91109.}}
\date{May 25, 1993}
\begin{document}
\maketitle
\begin{abstract}
It is believed that most quasars and galaxies present two common
features: the presence in their core of a supermassive object, and the
experience of one or more encounters with other galaxies. In this
scenario, it is likely that a substantial fraction of active galactic
nuclei harbour a supermassive binary, fueled by an accretion
disk. These binaries would certainly be among the strongest sources of
sinusoidal gravitational waves.
We investigate their evolution considering, simultaneously,
the accretion of the black hole's masses from the disk, and
the gravitational waves emitted during the orbital motion.
We also consider other astrophysical scenarios involving a
coalescing binary with non constant masses.
\end{abstract}

\section{Introduction}

Soon after the discovery of the astrophysical phenomena subsequently
indicated with the name of {\it Active Galactic Nuclei} (AGN), it was
argued that their power supply was ultimately gravitational in origin.
If this is the common feature of the wide range of models included in
the AGN category, then the natural conclusion is, as first pointed
out by Zeldovich \& Novikov (1964) and Salpeter (1964),
that an AGN `prime mover' is a
supermassive black hole (SBH). Among other compelling arguments in
favour of the black hole hypothesis are their efficiency and stability,
together with various observational discoveries, like rapid X-ray
variability, small scale radio jets, broad emission lines, etc.
(Blandford 1990; Osterbrock 1993).

For these reasons, most of the theoretical work about the AGN phenomenon
has been focused on the SBH hypothesis, namely that essentially all
active galaxies
contain $\sim 10^6-10^9\, \msun$ black holes in their nuclei, and that
these objects, together with their orbiting accretion disks, are the
prime movers for most of the powerful activity.

On the other hand, there are also compelling reasons to believe that a
great number of galaxies have undergone at least one merger since the
epoch of their formation (see Rees 1990, and references therein).
Indeed, many current models assume that the central object is activate,
or simply refueled, as a result of these interactions with another
galaxy (Osterbrock 1993).
This new evidence, along with the central
SBH hypothesis, suggests that, due to the dynamical friction exerted by
the surrounding environment, a certain number of active galaxies could
harbour a binary black hole system (BBHS), with separation of the order
of parsecs. This conclusion is supported by the observed bending
and apparent precession of radio jets emerging from AGN (Begelman,
Blandford, \& Rees 1980). In fact,
the S-symmetry observed in many radio sources and in a considerable
fraction of quasars at $z<1$ (Hutchings, Price, \& Gower 1988)
might be due to the presence
of such binaries. However, in many quasars the jets are strongly curved
on the milliarcsec scale as well. If this curvature is also due to
precession, much shorter precession periods, and thus smaller binary
separations, are required.
This indicates that mass flow into the galactic nucleus through the
accretion disk dominates not only the activity, but also the evolution
of the central system (Roos 1989).

Moving from these considerations, recent works have investigated the
evolution of a BBHS in the violently relaxed core of a merged galaxy,
taking into account the flow of gas and stars into the newly formed
nucleus (Begelman \al 1980; Roos 1989; Ebisuzaki, Makino, \&
Okumura 1991; Fukushige, Ebisuzaki, \& Makino 1992).

These AGN models are interesting per se, and also in conjunction with
the issue of gravitational waves (GW) detection. In fact, close
binary systems of supermassive compact objects are currently
considered as the most
certain observable sources for detectors like LIGO (Abramovici \al
1992) and VIRGO
(Bradaschia \al 1990),
and also for those experiments based on the Doppler tracking
of an interplanetary spacecraft (see, e.g., Thorne \& Braginsky 1976;
Bertotti \al 1992).
In the standard model for the emission of GW from a binary system, the
energy loss is proportional to the square of the third derivative of the
quadrupole moment. This description, however, does not take into
account other energy loss mechanism, like friction with an accretion
disk.

In this work, we analyze the evolution of a BBHS when the effects of
mass accretion and GW emission are considered
simultaneously. In \S 2 we will describe in detail a single BBHS, and
find, under
some reasonable assumptions,
the behavior of the separation between the two components. In \S 3
we show how the detection of these waves could provide some useful
information on the physical characteristic of the AGN, namely its
mass and accretion rate.
In \S 4 we extend our analysis to a simple population of BBHS
systems, and find the evolution law for the distribution function.
We conclude, in \S 5, pointing out other astrophysical situations in
which a relevant mass change can affect the evolution of the binary
system, and thus also the waveform of the emitted GW.

\section{Evolution of a binary system in presence of mass accretion}

The evolution of a close binary system in presence of mass loss or gain
has been thoroughly analyzed in the past, for example as a tool to
explain the irregularities and secular changes in some spectroscopic
binary stars (Kruszewski 1966).
Only later a similar analysis has been applied
to relativistic objects, like a neutron star binary (Clark \& Eardley
1977; Jaranowsky \& Krolak 1992).

We consider a binary system consisting of two supermassive black holes,
following circular newtonian orbits around the common
baricenter. This system, according to General Relativity, radiates
gravitational waves, which subtract energy and orbital angular
momentum from the system itself. The effect of this radiation is also to
circularize the orbit. For this reason we have assumed the orbits to be
circular. Also the frictional drag exerted by the surrounding gas and
stars during the formation and early evolutionary stage of the BBHS
should circularize the orbits, even if this point is still controversial
(Begelman \al 1980; Fukushige \al 1992).

The angular momentum carried away by the GW
in the unit of time is given by
the quadrupole formula (Peters 1964), which for point masses reads
(unit $c=G=1$ hereafter)
\begin{equation}
\frac{d J_{GW}}{dt}=\frac{32}{5}\frac{\mu^2 M_T^{5/2}}{a^{7/2}}\,,
\end{equation}
where $a$ is the separation between the BHs, $M_T=m_1+m_2$ is the total
mass, and $\mu=m_1 m_2/M_T$ is the reduced mass of the system.

When we consider such a system
in the core of an AGN, we must also take into account
the effect of the accretion
disk. This material, made up of gas, stars or even more
exotic objects, like ordinary black holes, is able to transfer or
subtract energy
and orbital angular momentum to/from the BBHS,
exerting a
frictional drag on the motion of the BHs. We will assume that i) the
mass falls radially, i.e.\ without any component tangential to the
orbits, ii) the direction of the drag force which acts on the BHs is
opposite to their velocity vectors, and, finally, iii) we neglect any
gravitational effect of the disk itself on the BBHS.

Then a simple newtonian analysis shows that
\begin{equation}
\frac{d}{dt}\left(\mu \sqrt{M a}\right) = -\frac{d J_{GW}}{dt}\,.
\end{equation}
For simplicity, we will assume, from now on, that the two BHs have the
same mass, $m$. Then, substituting \eq{1}
in \eq{2}, one gets the
following equation governing the separation $a$
\begin{equation}
\dot a = -\frac{128}{5}\left(m\over a\right)^3 - 3
\left(a\over m\right)\dot m\,.
\end{equation}
In order to integrate \eq{3} we should
have to know the time dependence
of $\dot m$. Unfortunately, while we can make some reasonable hypothesis
on the magnitude of $\dot m$, based on the Eddington assumption (see
later), very little can be said about the real time dependence of $\dot
m$, as well as the duration of the active phases. Due to our ignorance
on this subject, we will make the very simple assumption $\dot
m=\nu=const$. Then, integrating \eq{3} and taking $t=0$ as the
initial time, one gets
\newcommand{\fraz}{\frac{\tau_m}{16\tau_g}}
\begin{equation}
a(t)=a_0\left(1+\frac{t}{\tau_m}\right)\left[\left(1+\fraz\right)
\left(1+\frac{t}{\tau_m}\right)^{-16}-\fraz\right]^{1/4}\,,
\end{equation}
where we have introduced the two fundamental time scales
\begin{equation}
\tau_m\equiv \frac{m_0}{\nu}
\end{equation}
and
\begin{equation}
\tau_g\equiv\frac{5}{512}\frac{a_0^4}{m_0^3}\,.
\end{equation}
The meaning of these two quantities is as follows: $\tau_m$ gives the
mass e-folding time for accretion at the rate $\nu$, while $\tau_g$
gives the time required by the binary to coalesce as a consequence of
the energy and angular momentum carried away by the GW, assuming
circular orbits and $\nu=0$. For a very slow accretion rate, i.e.\ for
$\tau_m\gg\tau_g$, the time evolution of $a$ is dominated primarily
by GW emission,
and \eq{4} reduces in this limit to the well known formula
(Misner, Thorne, \& Wheeler 1973)
\begin{equation}
a(t)=a_0\left(1-\frac{t}{\tau_g}\right)^{1/4}\,.
\end{equation}

{}From \eq{4},
we can easily find the coalescing time, i.e.\ the time at
which $a$ vanishes,
\begin{equation}
\tau_c=\tau_m\left[\left(1+\frac{16 \tau_g}{\tau_m}
\right)^{1/16}-1\right]\,.
\end{equation}
In the limit $\tau_m\gg\tau_g$, by expanding \eq{8} in powers of
$\tau_g/\tau_m$, to first order we get
\begin{equation}
\tau_c=\tau_g\left[1-\frac{15}{2}\frac{\tau_g}{\tau_m}+
O\left(\tau_g\over\tau_m\right)^2\right]\,.
\end{equation}
Figure 1 shows the value of $\tau_c$ as a function of $\tau_m$, in units
$\tau_g=1$. As expected, the mass accretion speeds up the evolution fo
the system. In particular, if the two time scales are almost equal, then
the coalescing time is approximately  $20\%$ of $\tau_g$.

The crucial quantity which appears in equations (4) and
(8) is thus the ratio
\begin{equation}
\frac{\tau_g}{\tau_m}\simeq 3\times 10^6 \,m_8^{-4}
\left(a_0\over 1\um{pc} \right)^4
\left(\nu\over\msun/\um{yr}\right)\,,
\end{equation}
where $m_8$ is the initial mass in units of $10^8\msun$.
For values of $m$ and $\dot m$ typical of AGN models, this ratio
is usually bigger than one, unless one is willing to adopt extremely
small separations. This fact is of considerable importance, because we
are forced to abandon the usual idealized description, based on
equations (6) and (7),
in favour of the more general equations (4) and
(8). On
the other hand, the quantity given in \eq{10}
can not be arbitrarily large.
In the standard accreting model, the total radiated power is assumed to
be limited by the Eddington luminosity
\begin{equation}
L_E\simeq 2.6\times 10^{46}\, m_8 \:\um{erg}\um{s}^{-1}\,.
\end{equation}
In principle, this limit only applies for spherical accretion, when most
of the models
actually gives $L\ll L_E$ (Chang \& Ostriker 1985; Park \& Ostriker
1989), but valid
arguments suggest that this limit remains valid in every realistic
(i.e.\ without an unnatural segregation between radiation and fuel, see
Turner 1991) anisotropic model, which generally gives $L\minsim L_E$
(Rees 1984).

Associated with $L_E$ is an Eddington accretion rate, that would be able
to sustain an Eddington luminosity with efficiency $\epsilon$ for
conversion of mass into radiant energy
\begin{equation}
\dot m_E\approx \frac{0.2}{\epsilon}\, m_8\;\, \msun/\umtab{yr}\,.
\end{equation}
Equation (12) gives a lower limit for the e-folding time
\begin{equation}
\tau_E\simeq 4.4\times10^8\,\epsilon\:\um{yr}\,,
\end{equation}
which is independent of the mass. Note that \eq{12} implies an
exponential growth of the mass and its derivative, in contradiction with
our hypothesis $\dot m=const$. Nonetheless,
since we are implicitly assuming
that we observe the AGN for a time $T\ll\tau_m$, during which the
r.h.s.\ of \eq{12} can be assumed constant, we will
consider the
Eddington values (12) and (13)
as upper (lower) limits for $\dot m$ and
$\tau_m$, respectively. Thus, the ratio in \eq{10}
is bounded from above by the quantity
\begin{equation}
\frac{\tau_g}{\tau_E}\simeq\frac{7\times10^5}{\epsilon}\left(a_0\over
1\um{pc}\right)^4 m_8^{-3}\,.
\end{equation}
Some comments should refer to the unknown physical parameter $\epsilon$,
which describes the radiative efficiency. A detailed review of this
aspect has been given by Turner (1991), in the attempt of accounting for
the masses and luminosities of the oldest quasars ($z>4$). From his
considerations, one can adopt the reasonable value $\epsilon=0.1$,
independently of the details which perturb the fuel reservoir quite far
from the inner giant object.

Finally, another useful quantity is the critical separation, $a_c$,
defined as the initial separation which gives coalescence after a
certain time $t$. From \eq{8}, this is given by
\begin{equation}
a_c(t)=\left\{\frac{32 m_0^3 \tau_m}{5}\left[\left(1+
\frac{t}{\tau_m}\right)^{16}-1\right]\right\}^{1/4}\,.
\end{equation}
In using \eq{15},
we must remember that the origin of time was taken at
the instant of formation of the binary.
This quantity is of particular
importance when one is looking for the GW bursts resulting form the final
coalescences, since the probability of these events depends on the
percentage of systems which formed with separation around $a_c(t_{obs})$.

\section{The Braking Index}

In the previous section we have calculated the decreasing of the
separation $a$ as a consequence of the emission of GW and the accretion
of mass. Since the
frequency of the waves is twice the orbital angular frequency, the net
effect of their emission is an increase of their own frequency. If the
masses are constant, then $a$ is governed by \eq{7}, and from
Kepler's third law one finds
\begin{equation}
f(t)=f_0\left(1-\frac{t}{\tau_g}\right)^{-3/8}\,.
\end{equation}
However, in our case the situation is not so simple, since now the total
mass is increasing, and $a(t)$ is
given by the more complicate \eq{4}. The
overall effect is, thus, a more rapid increase of $f(t)$. This fact can
have considerable importance in relation to GW detection experiments. In
fact, a gravitational train emitted by a binary system is detectable as
a pure sinusoid only if the frequency remains in the same resolution
bin during the observations, i.e.\ as long as $\dot f T \minsim \Delta f
\equiv 1/T$, where $T$ is the duration of the experiment.
This constraint has been sometimes overlooked, in the past, since it
implies that the spectral region where the signal can be searched for
must be consistently restricted (Giampieri \& Tinto 1993).

In order to
describe the time evolution of $f(t)$, we take into account also its
derivatives $\dot f$ and $\ddot f$. In particular, the quantity of
interest is the {\it braking index}
\begin{equation}
k\equiv\frac{f \ddot f}{\dot f^2}.
\end{equation}
Now, Kepler's third law and \eq{4} give
\begin{equation}
\frac{\dot f}{f}=-\frac{3}{2}\frac{\dot a}{a} + \frac{1}{2} \frac{\dot
m}{m} = \frac{3 \cdot 2^{14/3}}{5} m^{5/3} (\pi f)^{8/3} + \frac{5\nu}{m}
\,.
\end{equation}
Deriving again \eq{18} with respect to $t$,
and neglecting the second
order terms in the nondimensional quantity $\nu (\pi m f)^{-8/3}$, we
eventually find
\begin{equation}
k\simeq \frac{11}{3}-\beta \nu (\pi m f)^{-8/3}\,,
\end{equation}
where $\beta$ is a numerical coefficient given by
$\beta\simeq 0.77$. The value $k=11/3$, corresponding to the limit case
$\nu=0$, is the constant value of the braking
index when one consider GW alone, as one can easily check from
\eq{16}.
If we were able to detect a sinusoidal GW at the frequency $f$ emitted
by a BBHS, then measuring $k$ one could determine the interesting
quantity $\nu m^{-8/3}$. Figure 2 shows this result when $\nu$ is
assumed to be equal to the Eddington limit (see eq.(12)),
with $\epsilon=0.1$.
Given the
GW frequency $f$, or, equivalently, the orbital period $P=2/f$,
and the deviation $\Delta k$ from the general
relativistic value $11/3$, one can determine the mass of the system.
Alternatively, if the mass is known, one can get relevant information on
$\nu$, as shown in Figure 3.

To be more useful, this method ought to be extended to more
realistic situations, where, for example, mass transfer between the two
components occurs. In other words, we would like to drop
the simplifying assumptions of equal masses and constant accretion
rates, and find the expression of $k$ which generalizes \eq{19}.
This
request can be easily fulfilled, since to obtain \eq{19}
we have not made
any use of the solution (4)
for $a(t)$; all we need is an expression for
$\dot a$ in terms of $a$ itself, $\mu, M_T$ and their derivatives
$\dot\mu,\dot M_T$.
{}From the adiabatic invariant one finds
\begin{equation}
\left(\dot a\over a\right)_m=-2\frac{\dot \mu}{\mu} - \frac{\dot
M_T}{M_T}\,,
\end{equation}
while \eq{1} gives
\begin{equation}
\left(\dot a\over a\right)_g=-\frac{64}{5} \frac{\mu M_T^2}{a^4}\,.
\end{equation}
Substituting equations (20) and (21) in
\eq{18}, we obtain, neglecting terms of order $O(\dot m/m)^2$ and
$O(\ddot m/m)$,
\begin{equation}
k\simeq
\frac{11}{3}-\frac{35}{96}\left(\frac{\dot \mu}{\mu} + \frac{2}{3}
\frac{\dot M_T}{M_T}\right) \mu^{-1} M_T^{-2/3} (\pi f)^{-8/3}\,,
\end{equation}
which reduces to \eq{19} when $m_1=m_2$ and $\dot m=const.$
{}From \eq{22} we see that
$k$, in the general case, can assume values above and below the
reference value $11/3$, according to the sign of the term
between parentheses in the last equation.
For example, if some material is flowing
from one BH (call it component `1') to the other one (component `2'),
then
\begin{eqnarray}
\dot \mu &=& \dot m_2\left(m_1-m_2\over M_T\right)\,,\\
\dot M_T &=& 0\,.
\end{eqnarray}
Thus, $k$ is bigger (lower) than $11/3$
if and only if
$m_1$ is less (more) massive than $m_2$. The mass transfer has, in
practical, no effect on $k$ as long as $m_1\simeq m_2$. Moreover, we
stress the possibility that a single BBHS could exploit different
behaviors of $\mu$ and $M_T$ during its evolution. From the
observational point of view, this fact gives rise to a dispersion of the
measured values of $k$, depending on which of the various possible
effects, namely mass loss-transfer-accretion, is dominant on each
particular system at that particular time. Thus, measuring $k$ in a wide
population of BBHS, one can get
very interesting information on the
variability of these systems, which seems to be
an important step toward the understanding of the fuel mechanisms.

\section{Evolutionary effects}

Following the discussion concluding the previous section, we will now
consider a population of BBHS.
Independently of their formation epoch and mechanism, we will start
considering them at a given `initial' time $t_0$, when each of them
is characterized by a separation $a_0$. For simplicity, we will assume
that $m_0$ and $\nu$ are equal in all these systems, and focus our
attention on the evolution of $a$. In
particular, we want to find the distribution function, i.e.\ the
function which gives the number of systems with a given separation $a$,
at an observation time $t$, given the same distribution at the initial
time $t_0$. To be more specific, we define the number of BBHS with
initial separation between $a_0$ and $a_0+da_0$ as $g_0(a_0) da_0/a_0$,
with $a_0$ belonging to the interval $[a_{\min},a_{\max}]$. The
lower limit $a_{\min}$ can be considered as the minimum separation
compatible with our newtonian assumption (for Post-Newtonian corrections
to the binary orbit see Lincoln \& Will 1990), while the upper
limit $a_{\max}$ is determined by the sensitivity and the bandwidth of
the receiver.

We want to find how $g_0$ evolves with time.
At a given time $t$, the binary
systems with $a_0<a_c(t)$ have disappeared, while those with $a>a_c(t)$
are distributed in accordance with the number conservation law, i.e.\
\begin{equation}
g(a,t)\frac{da}{a}=g_0(a_0)\frac{da_0}{a_0}\,,
\end{equation}
where $a$ is related to $a_0$ by \eq{4}.
Differentiating \eq{4} we thus find
\begin{equation}
g(a,t)=g_0(a_0)\underbrace{\left(a\over
a_0\right)^4}_{GW}
\underbrace{\left(1+\frac{t}{\tau_m}\right)^{12}}_{DISK}\;.
\end{equation}
In \eq{26}, we have made explicit the origin of the multiplicative
factors. In fact, the term $(a/a_0)^4$ is typically due to the emission
of GW alone, as can be seen differentiating \eq{7} (Bond \& Carr
1984).
Note, however, that in our case the relationship between $a$ and $a_0$
is not the same as in the `unperturbed' \eq{7}.
In other words, the
factorization in
\eq{26} is only apparent, since $\tau_m$ appears also,
through \eq{4},
in the term indicated with `GW'. Equation (26) can be
rewritten as
\begin{equation}
g(a,t)=g_0(a_0) F(a_0,t)\,,
\end{equation}
with the introduction of the transfer function $F(a_0,t)$, given
explicitly by
\begin{equation}
F(a_0,t)=1+\frac{\tau_m}{16\tau_g}\left[1-\left(1+
\frac{t}{\tau_m}\right)^{16}\right]\,.
\end{equation}
Note that this quantity is always less than one, and
depends on $a_0$ only through the GW time scale
$\tau_g$. Since $\tau_g\propto a_0^4$, this means that the evolution is
much faster near $a_{\min}$ than near $a_{\max}$. Figure 4 gives an
example of this behavior. Note also that, as expected,
$F(a_c(t),t)=0$.

\section{Conclusions}

In this work we have considered the effect of an accretion disk on the
GW emission from a BBHS. The main result is \eq{4},
which gives the
evolution of the separation $a$ as due to the GW emission together with
a (constant) mass accretion. We have found that the coalescing time can
be considerably shorter when the mass increases at the Eddington rate.
Finally, we gave an approximate expression for the braking index, in
terms of the reduced and total masses of the system.

We stress that most of the results
presented here have been obtained under quite general assumptions.
Therefore, they can be applied also in other interesting astrophysical
objects, like a binary pulsar, etc..
For example, we can consider a stellar binary system, loosing
its mass adiabatically. Then, assuming again equal masses and $\dot m =
const.$ ($\tau_m<0$ in this case),
we find that $a$ evolves according to the same \eq{4}.

In some circumstances, however,
the adiabatic hypothesis could appear inadequate.
In this case, \eq{4} is no longer valid. Anyway, we are able to
replace it, as long as the fractional change in the binding energy due
to the stationary mass loss is proportional to the fractional change of
the mass itself, i.e.
\begin{equation}
\frac{\delta E}{E}=\alpha \frac{\delta m}{m}\,.
\end{equation}
This case includes the previous one, since the adiabatic law corresponds
to $\alpha=5$, as well as other important cases; for example,
Jeans's mode of mass ejection implies $\alpha=3$ (Huang 1963).

Then \eq{3} must be replaced by
\begin{equation}
\dot a = -\frac{128}{5}\left(m\over a\right)^3-(\alpha-2)\left(a\over
m\right)\dot m\,,
\end{equation}
which can be easily integrated, to give
\begin{equation}
a=a_0\left(1+\frac{t}{\tau_m}\right)^{2-\alpha} F_\alpha(t)^{1/4}\,,
\end{equation}
where
\begin{equation}
F_\alpha(t)=\left\{\begin{array}{ll}
{\ds 1+\frac{\tau_m}{4(\alpha-1)\tau_g}\left[1-\left(1+\frac{t}{\tau_m}
\right)^{4(\alpha-1)}\right]} & \alpha\ne 1\\
\\
{\ds 1-\frac{\tau_m}{\tau_g}\ln\left(1+\frac{t}{\tau_m}\right)}
& \alpha=1
\end{array}
\right.
\end{equation}
Correspondingly, the coalescing time (8) becomes
\begin{equation}
\tau_c=\tau_m\cdot\left\{\begin{array}{ll}
{\ds \left(1+\frac{4(\alpha-1)\tau_g}{\tau_m}\right)^{1/4(\alpha-1)}-1}
& \mbox{\hspace{.4truecm}} \alpha\ne 1\\
\\
{\ds \exp(\tau_g/\tau_m)-1} & \mbox{\hspace{.4truecm}}\alpha=1
\end{array}
\right.
\end{equation}
When $|\tau_m|\gg\tau_g$, \eq{33} gives
\begin{equation}
\tau_c=\tau_g\left[1+\frac{5-4\alpha}{2}\frac{\tau_g}{\tau_m} +
O\left(\tau_g\over\tau_m\right)^2\right]\,.
\end{equation}
{}From \eq{34} we deduce that
\begin{eqnarray*}
\tau_c\magsim\tau_g & \mbox{for $\alpha\magsim 5/4$,}\\
\tau_c\simeq\tau_g & \mbox{for $\alpha\simeq 5/4$,}\\
\tau_c\minsim\tau_g & \mbox{for $\alpha\minsim 5/4$.}
\end{eqnarray*}

These formula could be encountered, for example, in a type
I Supernova progenitor, according to the Double Degenerate White Dwarfs
(DDWD) model (Iben \& Tutukov 1984).
In this model, both WD loose their H-rich envelope
just before the final coalescence. In the usual description of the DDWD
model the GW emission is considered only as a mechanism capable to get
the binary close enough for the common envelope phase (or phases) to
occur.
This idealized model, in which the various phenomena, namely the
mass transfer and ejection, and the final coalescence, take place with
different time scales, is obviously justified by the intention of giving
a satisfying description of the SNI progenitor. However, from the
results of our work, we can conclude that all these effects can
be taken into account simultaneously, and that they can
have considerable influence on the observability of this kind of
sources.

\section*{Acknowledgements}
This work was performed at JPL as a Resident Research Associate,
sponsored by the National Research Council (USA).

\newpage
\section*{References}
\begin{list}{}{\listparindent=0pt \parsep=0pt \leftmargin=0pt
\itemsep=8pt \itemindent=-8pt}

\paper{Abramovici, A., \al}{1992}{Science}{256}{325}
\paper{Begelman, M.C., Blandford, R.D., \& Rees, M.J.}{1980}
{Nature}{287}{307}
\paper{Bertotti, B., \al}{1992}{Astron.\ Astrophys.\ Suppl.\ Ser.}{92}
{431}
\proc{Blandford R.D.}{1990}{Active Galactic Nuclei: Saas-Fee Adv.\
Course 20}{T.Courvoisier \& M.Mayor}{Springer-Verlag, Berlin}
\paper{Bond, J.R., \& Carr, B.J.}{1984}{M.N.R.A.S.}{207}{585}
\paper{Bradaschia, C., \al}{1990}{Nucl.\ Instrum.\ Meth.\ A} {289}{518}
\paper{Chang, K.M., \& Ostriker, J.P.}{1985}{Ap.\ J.}{288}{428}
\paper{Clark, J.P.A., \& Eardley, D.M.}{1977}{Ap.\ J.}{215}{311}
\paper{Ebisuzaki, T., Makino, J., \& Okumura,
S.K.}{1991}{Nature}{352}{212}
\paper{Fukushige, T., Ebisuzaki, T., \& Makino, J.}{1992}{P.A.S.J.}
{44}{281}
\iom{Giampieri, G., \& Tinto, M.}{1993}{JPL Inter.\ Memo}{3394-93-049}
\paper{Huang, S.S.}{1963}{Ap.\ J.}{138}{471}
\paper{Hutchings, J.B., Price, R., \& Gower, A.C.}{1988}{Ap.\
J.}{329}{122}
\paper{Iben, I.Jr., \& Tutukov, A.V.}{1984}{Ap.\ J.\ Suppl.}{54}{335}
\paper{Jaranowski, P., \& Krolak, A.}{1992}{Ap.\ J.}{394}{586}
\paper{Kruszewski, A.}{1966}{Adv.\ Astron.\ Astrophys.}{4}{233}
\paper{Lincoln, C.W., \& Will, C.M.}{1990}{Phys.\ Rev.\ D}{42}{1123}
\book{Misner, C.W., Thorne, K.S., \& Wheeler, J.A.}{1973}{Gravitation}
{Freeman, San Francisco}
\paper{Osterbrock, D.E.}{1993}{Ap.\ J.}{404}{551}
\paper{Park, M.G., \& Ostriker, J.P.}{1989}{Ap.\ J.}{347}{679}
\paper{Peters, P.C.}{1964}{Phys.\ Rev.}{136}{1224}
\paper{Rees, M.J.}{1984}{A.R.A.\&A.}{22}{471}
\paper{---.}{1990}{Science}{247}{817}
\proc{Roos, N.}{1989}{Active Galactic Nuclei}{D.E.Osterbrock and
J.S.Miller}{Kluwer, Amsterdam}
\paper{Salpeter, E.E.}{1964}{Ap.\ J.}{140}{796}
\paper{Thorne, K.S., \& Braginsky, V.B.}{1976}{Ap.\ J.}{204}{L1}
\paper{Turner, E.L.}{1991}{Astron.\ J.}{101}{5}
\paper{Zel'dovich, Ya.B., \& Novikov, I.D.}{1964}{Sov.\ Phys.\ Dokl.}
{158}{811}
\end{list}

\newpage
\section*{Figure Captions}
\vspace{2truecm}
\begin{center}
{\bf Figure 1}
\end{center}

The coalescing time, $\tau_c$, as a function of the mass e-folding time,
$\tau_m$. Both quantities are expressed in units of the GW time scale,
$\tau_g$.

\vspace{1truecm}
\begin{center}
{\bf Figure 2}
\end{center}

How to determine the mass of the binary, knowing the GW
frequency $f$ and the braking index $k$, expressed here in terms of the
deviation from the GR value $11/3$. We have assumed an efficiency
$\epsilon=0.1$.

\vspace{1truecm}
\begin{center}
{\bf Figure 3}
\end{center}

How to determine the efficiency $\epsilon$, for a BH mass of $10^8\,
\msun$, from the knowledge of the GW frequency $f$
and the braking index $k$. As in fig.2, $\Delta k=11/3-k$.

\vspace{1truecm}
\begin{center}
{\bf Figure 4}
\end{center}

The evolution of the distribution function $g(a_0)$ at a generic time
$t$. Binaries with $a_0<a_c(t)$ have disappeared. Those with an initial
separation larger than $a_c(t)$ have evolved according to
\eq{27}. The initial distribution $g_0$ is assumed to be flat
over the interval $[0,a_{\max}]$ (dot-dashed line).

\end{document}